\documentstyle[12pt]{article}
\def\PRL #1 #2 #3{{\sl Phys. Rev. Lett.} {\bf#1} (#2) #3}
\def\NPB #1 #2 #3{{\sl Nucl. Phys.} {\bf B#1} (#2) #3}
\def\NPBFS #1 #2 #3 #4{{\sl Nucl. Phys.} {\bf B#2} [FS#1] (#3) #4}
\def\CMP #1 #2 #3{{\sl Commun. Math. Phys.} {\bf #1} (#2) #3}
\def\PRD #1 #2 #3{{\sl Phys. Rev.} {\bf D#1} (#2) #3}
\def\PLA #1 #2 #3{{\sl Phys. Lett.} {\bf #1A} (#2) #3}
\def\PLB #1 #2 #3{{\sl Phys. Lett.} {\bf #1B} (#2) #3}
\def\JMP #1 #2 #3{{\sl J. Math. Phys.} {\bf #1} (#2) #3}
\def\PTP #1 #2 #3{{\sl Prog. Theor. Phys.} {\bf #1} (#2) #3}
\def\SPTP #1 #2 #3{{\sl Suppl. Prog. Theor. Phys.} {\bf #1} (#2) #3}
\def\AoP #1 #2 #3{{\sl Ann. of Phys.} {\bf #1} (#2) #3}
\def\PNAS #1 #2 #3{{\sl Proc. Natl. Acad. Sci. USA} {\bf #1} (#2) #3}
\def\RMP #1 #2 #3{{\sl Rev. Mod. Phys.} {\bf #1} (#2) #3}
\def\PR #1 #2 #3{{\sl Phys. Reports} {\bf #1} (#2) #3}
\def\AoM #1 #2 #3{{\sl Ann. of Math.} {\bf #1} (#2) #3}
\def\UMN #1 #2 #3{{\sl Usp. Mat. Nauk} {\bf #1} (#2) #3}
\def\FAP #1 #2 #3{{\sl Funkt. Anal. Prilozheniya} {\bf #1} (#2) #3}
\def\FAaIA #1 #2 #3{{\sl Functional Analysis and Its Application} {\bf
#1} (#2) #3}
\def\BAMS #1 #2 #3{{\sl Bull. Am. Math. Soc.} {\bf #1} (#2)
#3} \def\TAMS #1 #2 #3{{\sl Trans. Am. Math. Soc.} {\bf #1} (#2) #3}
\def\InvM #1 #2 #3{{\sl Invent. Math.} {\bf #1} (#2) #3}
\def\LMP #1 #2 #3{{\sl Letters in Math. Phys.} {\bf #1} (#2) #3}
\def\IJMPA #1 #2 #3{{\sl Int. J. Mod. Phys.} {\bf A#1} (#2) #3}
\def\AdM #1 #2 #3{{\sl Advances in Math.} {\bf #1} (#2) #3}
\def\RMaP #1 #2 #3{{\sl Reports on Math. Phys.} {\bf #1} (#2) #3}
\def\IJM #1 #2 #3{{\sl Ill. J. Math.} {\bf #1} (#2) #3}
\def\APP #1 #2 #3{{\sl Acta Phys. Polon.} {\bf #1} (#2) #3}
\def\TMP #1 #2 #3{{\sl Theor. Mat. Phys.} {\bf #1} (#2) #3}
\def\JPA #1 #2 #3{{\sl J. Physics} {\bf A#1} (#2) #3}
\def\JSM #1 #2 #3{{\sl J. Soviet Math.} {\bf #1} (#2) #3}
\def\MPLA #1 #2 #3{{\sl Mod. Phys. Lett.} {\bf A#1} (#2) #3}
\def\JETP #1 #2 #3{{\sl Sov. Phys. JETP} {\bf #1} (#2) #3}
\def\JETPL #1 #2 #3{{\sl  Sov. Phys. JETP Lett.} {\bf #1} (#2) #3}
\def\PHSA #1 #2 #3{{\sl Physica} {\bf A#1} (#2) #3}
\def\CQG #1 #2 #3{{\sl Class. Quantum Grav.} {\bf #1} (#2) #3}
\def\SJNP #1 #2 #3{{\sl Sov. J. Nucl. Phys. (Yadern.Fiz.)} {\bf #1} (#2) #3}
\def\a{\alpha}\def\b{\beta}\def\d{\delta}\def\eps{\epsilon}

\def\L{\Lambda}
\def\Om{\Omega}
\def\Om{\Omega}
\def\cf{\cal F}
\def\be{\begin{equation}}\def\ee{\end{equation}}

\newcommand{\p}[1]{(\ref{#1})}
\begin{document}
\thispagestyle{empty}
\renewcommand{\thefootnote}{\fnsymbol{footnote}}
\begin{flushright}
HUB EP-98/28 \\
hep-th/9805110
\end{flushright}

\vspace{1truecm}
\begin{center}
{\large\bf Duality of Self--Dual Actions}

\vspace{0.4cm} Alexey Maznytsia$^1$, Christian R. Preitschopf$^2$
and
Dmitri Sorokin$^2$ {}$^3$
\footnote{Alexander von Humboldt Fellow \\
\phantom{mn} On leave from Kharkov Institute of Physics and
Technology, Kharkov, 310108, Ukraine}

\vspace{0.5cm}
$^1${\it Department of Physics and Technology,
Kharkov State University \\
310108, Kharkov, Ukraine\\
e--mail: alex$\_{}$maznytsia@hotmail.com
}

\vspace{0.5cm}
$^2${\it Humboldt-Universit\"at zu Berlin\\
Institut f\"ur Physik\\
Invalidenstrasse 110, D-10115 Berlin, Germany\\
e--mails: preitsch,sorokin@physik.hu-berlin.de}

\vspace{0.5cm}
$^3${\it Universit\`a Degli Studi Di Padova\\
Dipartimento Di Fisica ``Galileo Galilei''\\
ed INFN, Sezione Di Padova\\
Via F. Marzolo, 8, 35131 Padova, Italia}

\vspace{0.5cm}
{\bf Abstract}

\vspace{0.3cm}
\end{center}
Using examples of a $D=2$ chiral scalar and a duality-symmetric formulation of
$D=4$ Maxwell theory
we study duality properties of actions for describing chiral bosons.
In particular, in the $D=4$ case, upon performing
a duality transform of an auxiliary scalar field,
which ensures Lorentz covariance of the action,
we arrive at a new covariant
duality--symmetric Maxwell action, which contains a two-form potential
as an auxiliary field.  When the two-form field is gauge fixed
this action reduces to a
duality--symmetric action for Maxwell theory constructed by Zwanziger.
We consider properties of this new covariant 
action and discuss
its coupling to external dyonic sources. We also demonstrate that the
formulations considered are self--dual with respect to a dualization of
the field--strengths of the chiral fields.



\renewcommand{\thefootnote}{\arabic{footnote}}
\setcounter{footnote}0
\newpage

\renewcommand{\thefootnote}{\arabic{footnote}}
\setcounter{footnote}0
\section{Introduction}

Chiral bosons, or antisymmetric tensor fields with self--dual
field strengths, appear in
various theoretical models related to superstring theories, and reflect
the existence of a variety of important dualities connecting these theories
amongst each other.

Chiral field potentials are differential p--forms and, hence,
exist in a space whose dimension is
$D=2(p+1)$. If the metric signature of this space is Lorentzian, then
(anti-)self--duality requires the potential to be real, if $p$ is even,
or complex, if $p$ is odd. In the
latter case the theory can also be described by two real antisymmetric tensor
fields related by a duality condition.

The problem of a Lagrangian description of chiral bosons turned out
to be a complicated one, since manifest duality and space--time covariance
do not like to live in harmony with each other in one action.
Several non--covariant \cite{zwanziger}--\cite{ss} and space--time
covariant \cite{siegel}--\cite{pst} approaches
have been developed to solve this problem.

The first non--covariant duality--symmetric action was proposed by Zwanziger
for the description of $D=4$ Maxwell fields interacting with electrically and
magnetically charged particles \cite{zwanziger}.
In this formulation an
electromagnetic field is described by an $O(2)$ doublet of vector
potentials $A_m^{\a},~{\a}=1,2$, whose stress tensors
$F^{\a}_{mn}(A)=2{\partial}_{[m}A^{\a}_{n]}\equiv
{\partial}_{m}A^{\a}_{n}-{\partial}_{n}A^{\a}_{m}$
satisfy, in the absence of charged sources, the duality
condition
\be \label{D=4sd}
{\cf}^{\a}_{mn}(A)\equiv {\eps}^{\a \b}F^{\b}_{mn}(A)-
{1\over 2}{\eps}_{mnpq}F^{\a pq}(A)=0;\quad
{\cf}^{\a}_{mn}(A)=
{1\over 2}{\eps}^{{\a}{\b}}{\eps}_{mnpq}{\cf}^{{\b}pq}(A),
\ee
where ${\eps}^{\a \b}$ is the antisymmetric tensor 
(${\eps}^{12}=1$).

This condition follows from the Zwanziger action without charged sources
\be \label{freezwan}
S_Z=\int d^4x[-{1\over 8}F^{\a}_{mn}(A)F^{\a mn}(A)+
{1\over{4}}n^m{\cf}^{\a}_{mn}(A){\cf}^{\a np}(A)n_p],
\ee
which contains a constant normalized space--like vector $n^m$
$(n^mn_m=1)$ associated with a rigid Dirac string \cite{zwanziger}.
Its presence breaks the
$SO(1,3)$ Lorentz invariance of the action
\p{freezwan}  down to an $SO(1,2)$ group of
rotations in the hypersurface orthogonal to $n^m$.
Nevertheless, as we will show in this paper, the action \p{freezwan}
is invariant under non--manifest modified space--time transformations
which coincide with the $SO(1,3)$ Lorentz transformations on the mass shell.

To reduce the $A^\a_m$--field equations of motion derived from \p{freezwan}
to the duality condition \p{D=4sd} one should impose appropriate
boundary conditions on the field
strengths of the gauge fields \cite{zwanziger}.

Another non--covariant duality--symmetric action for Maxwell theory
was considered by Deser and Teitelboim \cite{deser}, 
and by Schwarz and Sen \cite{ss}. This action can be written in the 
following form:
\be \label{ss}
S=\int d^4x[-{1\over 8}F^{\a}_{mn}(A)F^{\a mn}(A)-
{1\over{4}}n^m{\cf}^{\a}_{mn}(A){\cf}^{\a np}(A)n_p].
\ee
It also contains a normalized constant space--like vector $n^m$
\footnote{In \cite{deser,ss} a {\it time}-like vector $n^m=(1,0,0,0)$ was used
to construct the action, but the choice of a constant {\it space}--like
vector is equally admissible.}.  We observe that the two actions
differ by a sign in front of the
second term. However,
they both produce the duality condition \p{D=4sd} and
describe a single Maxwell field. This sign difference results in
a difference in symmetry properties of the actions.
For instance, the duality condition arises as a consequence of
equations of motion, which follow from the action \p{ss}, upon
gauge fixing an additional local symmetry
\be \label{ss-symm}
\d A_m^{\a}=n_m{\Phi}^{\a}(x)
\ee
under which this action is invariant.
The Zwanziger action does not possess such
a symmetry.

A purpose of this paper is to establish the relationship between these two
actions. We will show that the Schwarz--Sen action and the Zwanziger action
are dual to each other. To perform the duality transform from one action
to another one should consider a manifestly covariant form of the 
action \p{ss} proposed in \cite{pst}, 
which contains an auxiliary scalar field $a(x)$
responsible for Lorentz covariance.

We analyze properties of the covariant formulation with respect to
the dualization of chiral boson fields and of the auxiliary scalar
field and find that the Zwanziger action appears upon the dualization
of the auxiliary field. In this way we get a new covariant form
of the duality--symmetric action from which the Zwanziger action is
obtained after gauge fixing a local symmetry. We also find a non-manifest
space--time invariance of the action \p{freezwan}.

The paper is organized as follows.

As an instructive simple example, in Section 2 we consider
a chiral scalar in two--dimensional space--time.
We demonstrate that its covariant action is self--dual with
respect to a dualization of the chiral scalar
as well as of the auxiliary scalar,
and show that in this case the actions of the Floreanini and Jackiw \cite{fj}
(or of a type \p{ss}) and of the Zwanziger type coincide.

In Section 3 we consider the same dualization procedure in application to
the duality--symmetric formulation of $D=4$ Maxwell theory. We find that the
dualization of the auxiliary scalar field $a(x)$ of this formulation into
a two--rank tensor field $B_{mn}(x)$
results in another duality--symmetric action.  We study
symmetry structure and dynamical equations of this new covariant action
and show that it reduces to the Zwanziger action \p{freezwan}
by gauge fixing its local symmetries.

In the Conclusion we discuss a problem of the duality relation between
the actions
\p{freezwan} and \p{ss} when they are coupled
to electrically and magnetically charged sources.

In the Appendix we demonstrate that chiral boson actions 
are self-dual with respect to a
dualization of chiral boson field strengths.

\section{Duality properties of the $D=2$ chiral
boson action}

We begin with the consideration of a covariant action for
a two--dimensional chiral scalar \cite{pstch1,pstch2}
\be \label{2pst}
S=\int d^2x[-{1\over 2}F_m(\phi )F^m(\phi )+
{1\over{2({\partial}_ra)({\partial}^ra)}}
({\partial}^ma{\cf}_m(\phi ))^2],
\ee
where $F_m(\phi )={\partial}_m\phi $ is the ``field strength" of the scalar
field $\phi (x) $, and ${\cf}_m(\phi )=F_m(\phi )-{\eps}_{mn}F^n(\phi )$
is its anti--self--dual part
$$
{\cf}_m(\phi )=-{\eps}_{mn}{\cf}^n(\phi ).
$$
To be space--time covariant the action \p{2pst} contains a normalized
derivative of an auxiliary scalar field $a(x)$.

    The action \p{2pst} is invariant under the following
transformations of the fields $a(x)$ and $\phi(x)$ \cite{pstch1,pstch2,cher}
\be \label{2asymm}
\d a=\varphi(x) ,\qquad
\d \phi ={{\varphi}\over{({\partial}a)^2}}
{\cf}^m(\phi ){\partial}_ma,
\ee
\be \label{2fsymm}
\d \phi =f(a),
\ee
with the parameters ${\varphi}(x)$ and $f(a)$, respectively.
Note that the latter is a function of $a(x)$.
Therefore, the transformations \p{2fsymm} are only
``semi--local", and there is no
first--class constraint associated with this symmetry in the covariant
formulation. The only first--class constraint is the one which generates
the transformations \p{2asymm} \cite{pstch2}.

The symmetry \p{2asymm} allows one to gauge away the auxiliary scalar
field $a(x)$ by fixing, for example, the temporal gauge
${\partial}_ma={\d}_m^0$.
Then the action \p{2pst} reduces to the non--covariant action
of Floreanini and Jackiw \cite{fj}, which is invariant under \p{2fsymm}
with $f(x^0)$. Now there appears a `weak' first--class constraint
associated with this constraint
but it rather corresponds to choosing boundary conditions for $\phi(x)$ than
eliminating a physical degree of freedom of the chiral boson (see, for
example \cite{neto}).

The $\phi(x)$ field equation, following from \p{2pst}, is
\be \label{2feqs}
{\eps}^{mn}{\partial}_n[
{1\over{2({\partial}a)^2}}
{\partial}_ma{\partial}_pa{\cf}^p(\phi )]=0.
\ee
Its general solution
\be \label{2fsol}
{1\over{2({\partial}a)^2}}
{\partial}_ma{\partial}_pa{\cf}^p(\phi )={\partial}_m{\zeta}
\ee
contains a scalar field $\zeta(x)$, which is supposed to be
a continuous and differentiable function of its arguments.
Projecting
\p{2fsol} on two
orthogonal vectors ${\partial}_ma$ and ${\eps}_{mn}{\partial}^na$, we get
\be \label{2fconstr}
{1\over 2}{\partial}_ma{\cf}^m(\phi )={\partial}_ma{\partial}^m{\zeta},
\ee
\be \label{2zetaconstr}
{\eps}^{mn}{\partial}_ma{\partial}_n{\zeta}=0.
\ee
The general solution to \p{2zetaconstr} can be expressed in terms of an
arbitrary function of $a(x)$
\be\label{zetsol}
\zeta=\hat f(a(x)).
\ee
We now notice that the left hand side
of \p{2fconstr} transforms under \p{2fsymm} as follows
$$
{\partial}_ma\d{\cf}^m(\phi )={\partial}_ma{\partial}^m{{f}(a)},
$$
Hence, we can use the transformations \p{2fsymm}
to put $\zeta=0$. Then \p{2fconstr} reduces to the anti--self--duality
(chirality) condition
\be\label{chir}
\partial_m{\cf}^m=\epsilon^{mn}\partial_m{\cf}_n=0 \quad \Rightarrow
\qquad {\cf}_m({\phi})=\partial_m\phi-\eps_{mn}\partial^n\phi=0.
\ee

\subsection{Dualization of the chiral scalar}
To study duality properties of the action \p{2pst} with respect to
a dualization of the field $\phi(x)$, we use a standard procedure.
We replace \p{2pst} with the following classically equivalent action
\be \label{2fparent}
S=\int d^2x[-{1\over 2}F_mF^m+
{1\over{2({\partial}_ra)({\partial}^ra)}}
({\partial}^ma{\cf}_m)^2+G^m(F_m-{\partial}_m{\phi})].
\ee
In addition to the fields $a(x)$ and $\phi(x)$ this action now contains
two independent auxiliary
vector fields $G^m(x)$ and $F^m(x)$, and ${\cf}_m\equiv F_m-{\eps}_{mn}F^n$.
The variation of \p{2fparent} with respect to $G^m$ gives the
expression for the field $F_m$ in terms of $\phi $
\be\label{Fphi}
F_m=\partial_m\phi,
\ee
which, when
substituted back into \p{2fparent}, yields the action \p{2pst}. If we
vary \p{2fparent} with respect to $F_m$ we obtain
the expression of $G^m$ in terms of $F^m$
\be\label{G}
G_m=-{2\over{({\partial}a)^2}}{\eps}_{mn}{\partial}^na
({\partial}_pa{\cf}^p)+{\eps}_{mn}F^n.
\ee
This can be inverted to express $F^m$ in terms of $G^m$
\be\label{F}
F_m={2\over{({\partial}a)^2}}{\partial}_ma
({\partial}_pa{\cal G}^p)+{\eps}_{mn}G^n,
\ee
$$
{\cal G}^m\equiv G^m-{\eps}^{mn}G_n.
$$
Note that on the mass shell \p{chir} $G_m$ and $F_m$ are dual to each other in
a usual sense, i.e. $G_m=\eps_{mn}F^n$.

Substituting \p{F} into the action \p{2fparent} we get
\be \label{2dualG}
S=\int d^2x[{1\over 2}G_mG^m+
{1\over{2({\partial}_ra)({\partial}^ra)}}
({\partial}^ma{\cal G}_m)^2+{\phi}({\partial}_mG^m)].
\ee

The variation of \p{2dualG} with respect to $\phi $ gives
\be\label{Gpsi}
{\partial}_mG^m=0 ~~\Rightarrow ~~ G^m={\eps}^{mn}{\partial}_n{\psi}
\equiv {\eps}^{mn}F_n(\psi ).
\ee
When the chirality condition \p{chir} is satisfied,
the expressions \p{G} and \p{Fphi} take the form
$$
G_m=\epsilon_{mn}F^n=\eps_{mn}\partial^n\phi(x),
$$
from which it follows that on the mass shell the scalar $\psi(x)$ of \p{Gpsi}
coincides with $\phi(x)$ up to a constant.
Substituting \p{Gpsi} into \p{2dualG} we
again recover the action \p{2pst}.

Thus, as one might expect, the action \p{2pst} is self--dual with respect to
a dualization of the chiral boson field $\phi$.

\subsection{Dualization of the auxiliary scalar}
Consider now properties of the action \p{2pst} under the
dualization of the auxiliary scalar field $a(x)$. In order to do this we
replace the action \p{2pst} with a classically equivalent action
\be \label{2parenta}
S=\int d^2x[-{1\over 2}F_m(\phi )F^m(\phi )+
{1\over{2u^ru_r}}
(u^m{\cf}_m(\phi ))^2+v^m(u_m-{\partial}_ma)],
\ee
where $F_m(\phi)=\partial_m\phi$, but $v^m(x)$ and $u^m(x)$ are now
independent auxiliary fields.
The equation of motion of the Lagrange multiplier $v^m$ gives
\be\label{u}
u_m={\partial}_ma,
\ee
while varying this
action with respect to $u^m$, we obtain the following expression for the field
$v^m$
\be \label{2vconstr}
v^m={1\over{(u^ru_r)^2}}{\eps}^{mn}u_n(u^p{\cf}_p(\phi ))^2.
\ee
Its consequences are
$$
v^mu_m=0,\qquad v^mv_m=-{1\over{(u^mu_m)^4}}
(u^p{\cf}_p(\phi ))^4,
$$
\be \label{2conseq}
{u^m\over{\sqrt{(u)^2}}}=
\eps^{mn}{v_n\over{\sqrt{-(v)^2}}}.
\ee
We see that normalized vectors $u_m$ and $v_m$ are dual to each other.

Substituting \p{2conseq} into the action \p{2parenta} we get
\be \label{2dualv}
S=\int d^2x[-{1\over 2}F_m(\phi )F^m(\phi )-
{1\over{2v^rv_r}}
(v^m{\cf}_m(\phi ))^2+a({\partial}_mv^m)].
\ee
Note that the second term of \p{2dualv} has the opposite
sign with respect to the
analogous term in \p{2parenta}. The same difference in the sign we have
observed when compared the Zwanziger action \p{freezwan} with the 
action \p{ss} for duality--symmetric Maxwell theory.

The equation of motion of the auxiliary field $a(x)$
$$
{\partial}_mv^m=0,
$$
which follows from \p{2dualv}, allows us to express $v_m$ as a derivative
of a scalar field
\be\label{v}
v^m={\eps}^{mn}{\partial}_nb(x).
\ee
Inserting this expression into
\p{2dualv} and taking into account the self--duality properties of
${\cf}_m(\phi)$ we get the action
\be \label{2dual}
S=\int d^2x[-{1\over 2}F_m(\phi )F^m(\phi )+
{1\over{2({\partial}_rb)({\partial}^rb)}}
({\partial}^mb{\cf}_m(\phi ))^2],
\ee
and find that it coincides with the action \p{2pst}
up to the replacement of
$a(x)$ with $b(x)$. Note that on the constraint surface \p{u}, \p{2conseq} and
\p{v} the scalar fields are related through the following condition
$$
{{\partial_ma}\over{\sqrt{({\partial}a)^2}}}=
{{\partial_mb}\over{\sqrt{({\partial}b)^2}}}
$$

We conclude that the action of two-dimensional chiral bosons is also
self--dual under the dualization of the scalar field $a(x)$.

\section{Dual descriptions of
duality--symmetric Maxwell theory}

      In this Section we study duality properties of a covariant
duality--symmetric action for  a free  Maxwell field \cite{pst}.
This action is the
covariant generalization of the action \p{ss} and has the form
\be \label{pst}
S=\int d^4x\big[-{1\over 8}F^{\a}_{mn}(A)F^{{\a}mn}(A)-
{1\over{4({\partial}^sa)({\partial}_sa)}}
{\partial}^ma{\cf}^{\a}_{mn}(A){\cf}^{{\a}np}(A){\partial}_pa\big].
\ee
The action \p{pst} is invariant under standard gauge transformations of
$A^{\a}_m$  \\ ($\d A^{\a}_m={\partial}_m\phi^\a(x)$), and unusual local
transformations
\be\label{unu1}
\d A^{\a}_m=({\partial}_ma){\Phi}^{\a},
\ee
\be\label{unu2}
\d a={\phi},\qquad
\d A^{\a}_m={{\phi}\over{({\partial}a)^2}}
{\eps}^{{\a}{\b}}{\cf}^{\b}_{mn}(A){\partial}^na,
\ee
(where ${\Phi}^{\a}(x)$ and ${\phi}(x)$ are corresponding parameters).
The first symmetry (eq. \p{unu1}) is the covariant analogue of the
symmetry \p{ss-symm} of the action \p{ss} \cite{ss}.
 It reduces the $A^\a(x)$--field equations of motion to the duality
relation between the field strengths $F^\a_{mn}(A)$
\be \label{duality}
{\cf}^{\a}_{mn}(A)=\eps^{\a\b}F^\b_{mn}(A)
-{1\over 2}\eps_{mnpq}F^{\a pq}(A)=0,
\ee
and the second symmetry \p{unu2}
allows one to gauge fix $\partial_ma(x)$ to be a constant time--like or
space--like vector, upon which the action reduces to \p{ss} (see
\cite{pst} for details).

As in the $D=2$ case, let us perform a duality transformation of the action
\p{pst} by dualizing the vector $u_m=\partial_m a(x)$. For this
replace the action \p{pst} with an equivalent action \footnote{An action of
this kind was discussed in \cite{pst}.}
\be\label{parent_on_a}
S=\int d^4x(-{1\over 8}F^{\a}_{mn}(A)F^{{\a}mn}(A)-
{1\over{4u^su_s}}u^m{\cf}^{\a}_{mn}(A){\cf}^{{\a}np}(A)u_p+
v^m(u_m-{\partial}_ma)),
\ee
where $u^m(x)$ is now an independent auxiliary vector field.
Variation of the action \p{parent_on_a} with respect to the Lagrange
multiplier $v^m$ gives the expression
\be \label{uex}
u_m={\partial}_ma,
\ee
which, after substitution into
\p{parent_on_a}, reduces it to the action \p{pst}.

The variation of the action with respect to $a(x)$ yields the relation
\be\label{b}
\partial_mv^m=0 ~~~\Rightarrow~~~
v^m=\eps^{mnpq}\partial_nB_{pq}.
\ee
We observe that, as in the $D=2$ case, the vectors $u_m$ and $v_m$
have a dual realization.
The first one is a vector ``field strength" of
the scalar field $a(x)$, while the second one
is the dual field strength of a two-form field $B_{mn}(x)$.

Finally, if we vary this action with respect to the field $u^m$, we obtain
the following equivalent expressions for $v^m$
\be \label{vex}
v^m=
-{1\over{2(u^su_s)^2}}u^mu^n{\cf}^\a_{np}(A){\cf}^{\a pq}(A)u_q
+{1\over{2(u^su_s)}}{\cf}^{\a mn}(A){\cf}^\a_{np}(A)u^p
\ee
\be\label{vex1}
v^m={1\over{2(u^su_s)^2}}\eps^{\a\b}\eps^{mnpq}u_n{\cf}^{\a}_{
p}{\cf}^{\b}_{ q},
\ee
where in \p{vex1} we have denoted with
${\cf}^\a_{mn}(A)u^n\equiv{\cf}^\a_m$ two vectors orthogonal to
$u^m$.

Eq. \p{vex1} is a kind of a duality relation between $v^m$ and $u_m$
which generalizes
eq. \p{2conseq} of the two--dimensional chiral--boson model of Section 2.
It has been obtained from \p{vex} by substituting ${\cf}^{\a mn}$ in the
second term of \p{vex} with its identical expression
\be\label{cfi}
{\cf}^{{\a}mn}(A)\equiv -{2\over{(u^su_s)}}u^{[m}{\cf}^{\a n]}-
{1\over{(u^su_s)}}\eps^{\a\b}\eps ^{mnpq}u_p{\cf}^{\b}_q.
\ee
  From \p{vex} and \p{vex1} it follows that
\be\label{vfu}
v^mu_m=0, \qquad v^m{\cf}^\a_{mn}(A)u^n\equiv v^m{\cf}^\a_m=0,
\ee
and
\be\label{vffv}
{1\over{(v^sv_s)}}v^n{\cf}^{\a}_{np}(A){\cf}^{{\a}pq}(A)v_q=-
{1\over{(u^su_s)}}u^n{\cf}^{\a}_{np}(A){\cf}^{{\a}pq}(A)u_q.
\ee
The latter is obtained by contracting \p{cfi} with $v^m$, taking
the square of both of its sides and using eqs. \p{vfu}.  Note that the
four vectors $u^n,~v^m$ and ${\cf}^{\a m}$ form an orthogonal basis in
$D=4$ space-time if ${\cf}^{\a m}$ are nonzero. This holds off the
mass shell (i.e. when eq. \p{duality} is not satisfied), so that the duality
transformation which we perform is an essentially
off--shell transformation, for
which the mass--shell condition is a singular point.
However, the action obtained
after this duality transition again yields the duality relation \p{duality}
between the two gauge potentials, as we shall see below.

Substituting \p{vffv} into \p{parent_on_a}, we get the
following action dual to \p{pst}
\be \label{zwanziger}
S=\int d^4x(-{1\over 8}F^\a_{mn}(A)F^{\a mn}(A)+
{1\over{4v^sv_s}}v^m{\cf}^\a_{mn}(A){\cf}^{\a np}(A)v_p).
 \ee
Its second term has an opposite sign to that in \p{pst} and
contains the auxiliary field $v^m=\eps^{mnpq}\partial_nB_{pq}$ (see \p{b}),
which can be gauge fixed to a constant vector
by use of a local symmetry of \p{zwanziger}. After this gauge fixing
the action \p{zwanziger} reduces to the Zwanziger action \p{freezwan}.
To convince oneself that such a symmetry does exist, perform
a general variation of the action \p{zwanziger} with respect to the
independent fields $A^\a_m$ and $B_{mn}$. This variation can be written in
the following form
\be \label{variation}
\d S=\int
d^4x\big[2(\d A^\a_n -{2\over{v^2}}\d B_{nq}{\cf}^{\a qr}(A)v_r){\eps}^{\a\b}
{\partial}_m({1\over{v^2}}v^{[m}{\cf}^{n]p\b}(A)v_p)
\ee
$$
+ {2\over{v^2}}\d B_{mn}v^m{\cf}^{\a lp}(A)v_p \eps^{\a\b}
\partial_l({1\over{v^2}}{\cf}^{\b nq}(A)v_q)\big].
$$
The variation \p{variation} vanishes when
$$
\d B_{mn}=-{1\over 2}\eps_{mnpq}v^p\L^q ~~~\Rightarrow
~~~\d v^m=2{\partial}_n(v^{[m}{\L}^{n]})=\L^n\partial_nv^m-v^n\partial_n\L^m-
v^m\partial_n\L^n,
$$
\be \label{vsymm}
\d A^{m\a}=-{1\over{v^sv_s}}
{\eps}^{mnpq}v_n{\L}_p{\cf}^\a_{qt}(A)v^t,
\ee
and
\be \label{Bsymm}
\d B_{mn}=2{\partial}_{[m}C_{n]},
\ee
where ${\L}^m(x)$ and $C^m(x)$ are vector parameters of these
transformations.
When $\L_m$ in \p{vsymm} is proportional to $v_m$, the variation of
$B_{mn}$ becomes trivial.  Therefore, only three independent
components of ${\L}^m(x)$ participate in \p{vsymm}.

The form of the transformation of $v^m$ in \p{vsymm} is reminiscent of general
coordinate transformations of vectors in curved space--time and differs from
the latter only by the presence of the last term, which looks like a dilatation.
Thus, by use of \p{vsymm}
and \p{Bsymm} we can gauge the normalized vector
${{v^m}\over{\sqrt{v^rv_r}}}$ to a unit constant space-like vector $n^m$
\footnote{Notice that $B_{mn}$ enters the action \p{zwanziger} through the
normalized vector field strength ${{v^m}\over{\sqrt{v^rv_r}}}$ which has
three independent components.}.  After such a gauge fixing the off--shell
model loses the standard manifest Lorentz invariance. The latter acquires
a modified form of a combination of the standard Lorentz transformations
with the transformation \p{vsymm} whose parameter is chosen in such a way
that these combined transformations do not affect the constant vector
$n^m$.

Standard Lorentz transformations of the 4--vector
$n^m$ are
\begin{equation}\label{lo}
\d _L n^m={\Omega}^{m}_{~~n}n^n,
\end{equation}
where $\Omega_{mn}$ is an antisymmetric infinitesimal Lorentz parameter.
At the same time we transform the
vector $n^m$ as in \p{vsymm}:
\begin{equation}\label{la}
\d _{\L}n^m=(n^q{\partial_q})(n^mn^p{\L}_p-{\L}^m).
\end{equation}
(on the r.h.s. of this relation the gauge fixing condition
${{v^m}\over{\sqrt{v^rv_r}}}=n^m$ has been taken into account).
Using this transformation we can compensate the Lorentz rotation \p{lo}
of $n^{m}$
if the parameter $\L_m$ is chosen to be
$
{\L}_m={\Omega}_{mn}x^n.
$
Then
$$
(\d _{L}+\delta_\L )n^m=0.
$$
But this combined transformation acts non-trivially on the fields $A^{\a}_m$.
As a result the
modified Lorentz transformation of $A^\a_m$ takes
the following form
\be\label{modlo}
\d
A^{\a}_m={\Om}_m^{~~n}A_n^{\a}+{\Om}^{pq}(x_{p}{\partial}_{q})A_m^{\a}
-{\eps}_{mnpq}n^n{\Om}^{pr}x_r{\cf}^{{\a}qt}(A)n_t,
\ee
The transformations \p{modlo} reduce to the standard Lorentz
transformations on the mass shell, when their last term (proportional to
${\cf}^{\a}_{mn}(A)$) vanishes.  This term appeared due to the
contribution of the transformations \p{vsymm}
with the parameter ${\L}^m={\Om}^{mn}x_n$.

Substituting the gauge condition
${{v^m}\over{\sqrt{v^rv_r}}}=n^m$ back into the covariant action
\p{zwanziger}  we reduce it to the free Zwanziger action \p{freezwan}.
Thus the Zwanziger action possesses a non--manifest space--time symmetry
\p{modlo},
which one may regard as a dual analogue of the corresponding space--time
symmetry of the non-covariant duality symmetric actions proposed in
\cite{Henn,ss}.

 From the general variation \p{variation} of the action we derive
the equations of motion of $A^\a_m$ and $B_{mn}$
\be\label{A}
{\eps}^{\a\b}{\partial}_m({1\over{v^2}}v^{[m}{\cf}^{n]p\b}(A)v_p)=0,
\ee
\be\label{B}
\eps^{\a\b}v^{[n}\partial_l({1\over{v^2}}{\cf}^{m]r\b}v_r)
{\cf}^{\a lp}(A)v_p=0.
\ee

The second--order differential equation \p{A}
can be reduced to the first--order duality condition \p{duality}
\footnote{Note that the $B_{mn}(x)$--field equation
\p{B} is automatically satisfied
when the duality condition holds, but if
we do not reduce $A_m(x)$--field equations to the duality condition then
\p{B} may impose additional restrictions on the form of
${\cf}^\a_{mn}(A)$, which would be of interest to analyze. This situation
is novel in comparison with equations of motion of $A^{\a}_m(x)$ and
$a(x)$ obtained from the action \p{pst}. In that case the $a(x)$--field
equation was a direct consequence of the $A^{\a}_m(x)$--field equations.}
by choosing appropriate boundary conditions for the gauge field strengths,
as in \cite{zwanziger}, or
equivalently by using the following semi--local
symmetry of the action \p{zwanziger}
\be \label{quasi}
\d A_m^{\a}=
\Phi^\a_m(x)-{1\over{v^2}}v_m(v^n\Phi^{\a}_n).\quad
\ee
The parameters ${\Phi}^{{\a}}_m$ in \p{quasi} are restricted to satisfy
\be \label{rest}
(v^p\partial_p)(\Phi^\a_m(x)-{1\over{v^2}}v_m(v^n\Phi^{\a}_n))
+(\partial_mv^p)(\Phi^\a_p(x)-{1\over{v^2}}v_p(v^n\Phi^{\a}_n))=0.
\ee
We have not been able to solve eq. \p{rest} in terms of unrestricted fields
in a covariant way, but when the symmetry
\p{vsymm} is fixed by the gauge ${{v^m}\over{\sqrt{v^2}}}=n^m$,
a solution can be obtained.
Then \p{quasi} and \p{rest} reduce to
\be\label{quasi1}
\d A_m^{\a}=\Phi^\a_m(x)-n_m(n^l\Phi^{\a}_l),
\ee
\be\label{rest1}
n^l\partial_l\Phi^\a_m(x)=0.
\ee
The latter is easily solved in terms of functions of three independent
arguments
\be\label{3}
\Phi^\a_m=\Phi^\a_m(y^l), \quad y^l=x^l-n^l(n_px^p), \quad
y^ln_l\equiv 0.
\ee
This symmetry of the free Zwanziger action
\p{freezwan} is an analogue of the symmetry \p{2fsymm} of the $D=2$
chiral boson model in a non-covariant gauge when, for example,
$a(x)=n^m{\eps}_{mn}x^n$.
Note that the coordinates $x^l-n^l(n_px^p)$ coincide
with gauge fixed components of the vector
$Y^l={1\over{\sqrt{v^2}}}\epsilon^{lmnp}v_mB_{np}$, when
${{v^m}\over{\sqrt{v^2}}}=n^m$ and $B_{mn}=\eps_{mnpq}n^px^q$.

The transformations \p{quasi}--\p{3} can be used to
reduce
eq. \p{A} to the duality condition \p{duality}. The procedure is analogous
to the one considered
in the previous Section to get the $D=2$ chiral boson equation.

Note that upon gauge fixing $v^m$ eqs. \p{A} can be rewritten
as Maxwell equations for a single field strength $F_{mn}$. 
(This observation was a starting point of Zwanziger
for the construction of the action \cite{zwanziger}):
$$
\partial_mF^{mn}=0,\qquad \partial_mF^{*mn}=0,
$$
where
$$
F_{mn}\equiv -2n^{[m}F^{1n]p}n_p + {\eps}^{mnpq}n_pF^2_{qr}n^r,
$$
$$
F^{*mn}={1\over 2}{\eps}^{mnpq}F_{pq}=-2(n^{[m}F^{2n]p}n_p)-
{\eps}^{mnpq}n_pF^1_{qr}n^r.
$$
When the boundary conditions are chosen such that $F^1_{mn}=-F^{2*}_{mn}$
(as in eq. \p{duality}), the field strength $F_{mn}$ coincides
with $F^1_{mn}$. 

We have seen that the parameters of the transformations \p{quasi}--\p{3}
depend only on three of the four space--time coordinates.
Therefore, they do not reduce the number of four--dimensional physical
degrees of
freedom, and their fixing is equivalent to choosing
appropriate boundary conditions on the fields of the model on the
3--dimensional hypersurface, as it was done in \cite{zwanziger}.
This is in contrast to the duality--symmetric version
of Maxwell theory based on the action \p{pst}. The transformations \p{unu1}
form a full--fledged local symmetry of \p{pst},
they yield first--class constraints \cite{pstch2} and eliminate degrees of
freedom of $A^\a_m(x)$ reducing them to that of a single
Maxwell field \cite{ss,pst}.
This example teaches us that dual formulations of one and the same model
may have very different symmetry structure.

One can also show that, as the $D=2$ action
\p{2pst}, the actions \p{pst} and  \p{zwanziger} are self--dual with respect
to a dualization of the gauge field strengths. The
proof is presented in the Appendix. This self--duality is quite natural and
reflects a basic property of the chiral boson actions in any
even space--time dimension \cite{Henn,ss}.

\section{Conclusion and discussion}
By the use of examples of a $D=2$ chiral scalar and $D=4$
Maxwell theory we have
studied duality properties of a Lorentz--covariant action for chiral bosons
\cite{pst}.

The action was shown to be self--dual with respect to a dualization of
the field strengths of the chiral fields.

The dualization of the auxiliary field (whose role in the action is to
preserve
Lorentz invariance) resulted in another form of the covariant
duality--symmetric action, where the role of the auxiliary field is taken
by a $(D-2)$--rank tensor field. This action has symmetry properties
which
are different from that of its dual partner.
Even though only the $D=2$ and $D=4$ case have been considered in detail,
we have checked that dual actions for chiral bosons of this kind exist 
in any even space--time dimension.

In the $D=4$ case it has been  shown that the dual covariant
action \p{zwanziger}
for a Maxwell field possesses a local symmetry which allows one to gauge fix
the vector field strength of the auxiliary field $B_{mn}(x)$ to be a constant
vector, and this reduces the action to the non--covariant Zwanziger action.
As a consequence, we have established the duality relationship between the
two duality--symmetric actions for free
Maxwell theory.

We have also found a hidden space--time invariance of the Zwanziger action
which
on the mass shell becomes a conventional Lorentz invariance.

A problem which requires further study is the relationship between the two
actions when they are coupled to dyonic sources.

Consistent coupling of the vector fields $A^\a_m(x)$ in the action
\p{ss} (and its covariant generalization \p{pst})
to charged currents $j^\a_m(x)$
requires incorporation of non--local terms which can be
interpreted as a contribution of a Dirac string attached to magnetically
charged objects \cite{dirac,deser_s,bm}.
In the formulations under
consideration the introduction of such terms is a requirement of local
symmetries \p{unu1} and \p{unu2}
which must be preserved when the coupling is made \cite{bm}.

Consider this in more detail. To generalize the action \p{pst} to one which
describes a coupling of $A^\a_m(x)$ to currents $j^\a_m(x)$ we replace
the field strengths of the
gauge fields in \p{pst} with modified stress tensors
\be \label{modten}
F^{\a mn}(A)~~~\rightarrow~~~\hat F^{\a mn}(A)=F^{\a mn}(A)+{1\over 2}
{\eps}^{\a\b}{\eps}^{mnpq}{S}^{\b}_{pq}
\ee
containing a tensor field ${S}^{\a}_{mn}$ determined by the condition
\be \label{nonloc_eq}
{\partial}^m{S}^{\a}_{mn}+j^{\a}_n=0,
\ee
and add to the action the minimal interaction term.
As a result we obtain the following action
\footnote{Here and below the vector fields $v^m$ and $u^m$ are defined as
in the previous Section (see \p{uex} and \p{b}).}
\be \label{spst}
S=\int d^4x(-{1\over 8}\hat F^{\a}_{mn}\hat
F^{{\a}mn}- {1\over{4u^su_s}}u^m \hat{\cf}^{\a}_{mn}\hat{\cf}^{{\a}np}u_p
-{1\over 2}j^{m\a}A^\a_m)
\ee
($\hat{\cf }^{\a mn}\equiv {\eps}^{\a \b}{\hat F}^{\b mn}-
{1\over 2}{\eps}^{mnpq}\hat F^{\a}_{pq}$).
One may check
that this action possesses the symmetries \p{unu1} and \p{unu2}
if we replace all $F^{\a}_{mn}(A)$
with $\hat F^{\a}_{mn}(A)$ in the transformation laws. Using these
symmetries we can obtain the modified duality condition
\be\label{mod}
\hat {\cf}^{\a}_{mn}=0 \quad \Rightarrow \quad
\hat F^1_{mn}=-{1\over 2}\eps_{mnpq}\hat F^{2pq}
\ee
as a solution to the equations of motion of this model.
Differentiating \p{mod} and taking into account \p{nonloc_eq} we get
the Maxwell equations with electric and magnetic currents
\be\label{max}
\partial_m\hat F^{1mn}=j^{1n}, \qquad \partial_m{~}^*\hat F^{1mn}=j^{2n}.
\ee

Dualizing the action \p{spst} we arrive at a generalization of the action
\p{zwanziger}, which has the same form as \p{spst} but with $v^m(x)$ instead
of $u^m(x)$ and with the plus sign in front of the second term. This Lorentz
covariant action is invariant under transformations \p{vsymm} and \p{quasi}
with
modified field strengths \p{modten}, and yields eqs. \p{mod} and \p{max} as a
consequence of equations of motion and gauge fixing.
When the auxiliary vector is fixed to be a constant the action reduces to a
non--covariant action, which differs in the non--local coupling terms from the
original Zwanziger action \cite{zwanziger} which also
describes an electromagnetic
interaction of dyons. The latter contains,
in addition to the action \p{freezwan}, only the minimal
coupling term $\int d^4x A^\a_m(x)j^{\a m}$. This Zwanziger action
is not invariant under the transformations \p{modlo} and \p{quasi1}, and
eqs. \p{mod}, \p{max} are singled out as particular solutions of
$A^\a_m$--field
equations by choosing appropriate boundary conditions \cite{zwanziger}.

We have now seen that for the classical duality--symmetric action of chiral
bosons coupled to charged sources to be manifestly or non--manifestly
space--time invariant the electric--magnetic coupling should be non--local.
In spite of this fact, one may expect that the quantization of these actions
will give the same result as Zwanziger quantization \cite{zwanziger}, since
the space of quantized physical modes is one and the same in both cases.

\bigskip
\bigskip
\noindent
{\bf Acknowledgements} 
D.S. is grateful to Paolo Pasti and Mario Tonin for useful discussions.
The work of A.M. and D.S. was partially supported by the 
INTAS Grants N 93--493--ext and N 97--0308. D.S. also thanks the Alexander
von Humboldt Foundation for providing him with a European Research Grant
for visiting Padua University where this work was completed.

\section{Appendix. Self--duality of chiral boson actions}

Consider the following action, equivalent to
\p{pst}
\be \label{parent_on_F}
S=\int d^4x(-{1\over 8}F^{\a}_{mn}F^{{\a}mn}-
{1\over{4({\partial}^sa)({\partial}_sa)}}
{\partial}^ma{\cf}^{\a}_{mn}{\cf}^{{\a}np}{\partial}_pa+
{1\over 4}G^{\a}_{mn}(F^{{\a}mn}-2{\partial}^{[m}A^{n]{\a}})),
\ee
containing now antisymmetric tensor fields $F^{\a}_{mn}$ and
$G^{\a}_{mn}$, two vector fields $A^{\a}_m$ and one scalar field $a(x)$ as
independent ones.

Variation of the action \p{parent_on_F} with respect to the Lagrange
multiplier $G^{\a}_{mn}$ yields the expression for
$F^{\a}_{mn}$ as the field strengths of the vector fields $A^{\a}_m$. If
we substitute this relation back into the action \p{parent_on_F}, the
latter reduces to \p{pst}.

To get the action dual to \p{pst} with respect to $F^\a_{mn}$ and $G^\a_{mn}$
consider the $F^{\a}_{mn}$--variation of \p{parent_on_F}. This variation
yields a duality--like relation
\be \label{Gconstr}
G^{{\a}mn}=
-{2\over{({\partial}a)^2}}
{\epsilon}^{mnpq}{\partial}_pa{\cf}^{\a}_{qr}{\partial}^ra-
{1\over 2}{\eps}^{{\a}{\b}}{\eps}^{mnpq}F^{\b}_{pq},
\ee
which  expresses the fields $G^\a_{mn}$ through
$F^{\a}_{mn}$
and becomes the standard duality relation on the mass shell \p{duality}.
By performing algebraic manipulations we can invert this relation such
that it will
express $F^{\a}_{mn}$ in terms of  $G^{\a}_{mn}$
\be \label{Fconstr}
F^a_{mn}= {{4}\over{({\partial}a)^2}}
{\eps}^{{\a}{\b}}{\partial}^{[m}a{\cal G}^{n]p{\b}}{\partial}_pa-
{1\over 2}{\eps}^{{\a}{\b}}{\eps}^{mnpq}G^{\b}_{pq},
\ee
where ${\cal G}^{\a}_{mn}={\eps}^{{\a}{\b}}G^{\b}_{mn}-{1\over 2}
{\eps}^{mnpq}G^{\a}_{pq}$.

If we vary the action \p{parent_on_F} with respect to $A^\a_m$, we obtain the
equations
$$
{\partial}_mG^{\a mn}=0,
$$
which can be regarded as Bianchi identities for a dual field strength,
whose general solution can be written in terms of two
vector fields $B^\a_m$
\be \label{Gsolution}
G^{{\a}mn}(B)=-{\eps}^{\a\b}{\eps}^{mnpq}{\partial}_pB^\b_q
\equiv-{1\over 2}{\eps}^{{\a}{\b}}{\eps}^{mnpq}F^\b_{pq}(B).
\ee
(Note that on the mass shell the $A^\a_m$ and $B^\b_m$ fields coincide).
Inserting this expression for $G^{\a}_{mn}$ into \p{Fconstr}
and then plugging the latter into the action \p{parent_on_F}, we recover
\p{pst}, where the gauge fields are $B^\a_{m}$
\be \label{pst(B)} 
S=\int d^4x[-{1\over 8}F^{\a}_{mn}(B)F^{{\a}mn}(B)-
{1\over{4({\partial}^sa)({\partial}_sa)}}
{\partial}^ma{\cf}^{\a}_{mn}(B){\cf}^{{\a}np}(B){\partial}_pa].
\ee

Thus, we conclude that, as one should expect, the action \p{pst} is self--dual
under the dualization of $A^\a_m$.

In the same way one can show that chiral boson actions in any dimension
$D=2(p+1)$ are self--dual, which in fact is their basic property.

\end{document}